\documentclass[12pt,letterpaper]{article}

\usepackage{graphics}
\usepackage{achemso}

\begin{document}

\title{\bf Temperature dependence of evaporation coefficient of water
in air and nitrogen under atmospheric pressure; study in water
droplets}
\author{{\bf M Zientara, D Jakubczyk\thanks{jakub@ifpan.edu.pl}, K Kolwas and M
Kolwas}\\
Institute of Physics of the Polish Academy of Sciences\\
Al.Lotnik\'{o}w 32/46, 02-668 Warsaw, Poland}

\maketitle

\begin{abstract}
The evaporation coefficients of water in air and nitrogen were
found as a function of temperature, by studying the evaporation of
pure water droplet. The droplet was levitated in an electrodynamic
trap placed in a climatic chamber maintaining atmospheric
pressure. Droplet radius evolution and evaporation dynamics were
studied with high precision by analyzing the angle-resolved light
scattering Mie interference patterns. A model of quasi-stationary
droplet evolution, accounting for the kinetic effects near the
droplet surface was applied. In particular, the effect of thermal
effusion (a short range analogue of thermal diffusion) was
discussed and accounted for. The evaporation coefficient $\alpha $
in air and in nitrogen were found equal. $\alpha $ was found to
decrease from $\sim 0.18$ to $\sim 0.13$ for the temperature range
from 273.1 K to 293.1 K and follow the trend given by Arrhenius
formula. The agreement with condensation coefficient values
obtained with essentially different method by Li et al. \cite{Li}
was found excellent. The comparison of experimental conditions
used in both methods revealed no dependence of
evaporation/condensation coefficient upon droplet charge nor
ambient gas pressure within experimental parameters range. The
average value of thermal accommodation coefficient over the same
temperature range was found to be $1\pm 0.05$.
\end{abstract}


\section{Introduction}

Many problems of science and technology are related to the
evaporation from droplets and condensation on them. Cloud and
aerosol microphysics together with construction of climate models
\cite{McFiggans,Laaksonen,Ackerman}, electrospraying \cite{Grimm},
combustion \cite{Sazhin}, jet printing (compare \cite{Percin}) and
spray painting (compare \cite{Sommerfeld}) are just some areas of
relevance. Though they concern large sets of coexisting droplets,
the understanding of transport processes at the surface of a
single droplet is vital for solving them properly. Mass and heat
transport processes at (nearly) flat interface can be efficiently
modeled as a diffusion phenomenon. However, the evolution of small
droplets is significantly influenced by effusion, which takes
place in effectively collision-free region in the very vicinity of
the interface (up to the mean free path of surrounding gas
molecules). In order to account for this phenomenon, a so called
evaporation (condensation) or mass accommodation coefficient
$\alpha $ is introduced besides the diffusion coefficient.
Likewise the thermal conductivity coefficient should be
accompanied by a thermal accommodation coefficient $\alpha _{T}$.
These coefficients describe transport properties of the liquid-gas
interface. The mass accommodation coefficient can be perceived as
the probability that a molecule (e.g. water) impinging on the
interface from the gaseous phase side enters into the liquid phase
and does not rebound. Analogically, the thermal accommodation
coefficient determines the probability that a molecule impinging
on the interface attains thermal equilibrium with the medium on
the opposite side. The considerations of evaporation and
condensation coefficients are considered equivalent and the values
of these coefficients - equal \cite{Pruppacher}. Both $\alpha $
and $\alpha _T$ coefficients are phenomenological and should
describe only the properties of the very interface. All other
processes influencing mass and heat transport, such as chemistry
of the interface or the electrostatic interactions should be
accounted for separately \cite{Shi}. It is agreed, however, that
$\alpha $ might possibly exhibit some temperature dependence
\cite{Pruppacher,Marek}.

Many attempts have been made over nearly a century, to determine experimentally the
values of $\alpha $ and $\alpha _{T}$ for water, but the results obtained by
different authors spanned from $\sim 0.001$ to 1 for $\alpha $ and from $\sim 0.5$
to 1 for $\alpha _{T}$ (see e.g.
\cite{Li,Winkler,Hagen,Zagaynow,Sageev,Gollub,FukutaMyers,Shaw,Xue} and
\cite{Pruppacher,Marek,4x} for revues). A variety of experimental methods was used.
Both condensation on and evaporation from the surface of bulk liquid, liquid films,
jets, and droplets were investigated in various environments (vacuum, standard air,
passive or reactive atmospheres) under various pressures and for various water vapor
saturations. Small droplets, such as encountered in clouds, have been favored, since
kinetic effects manifest strongly for them. Suspended droplets, trains of droplets,
clouds of droplets and single trapped droplets were studied.

We must admit that in our studies we have also experienced the flow of kinetic
coefficients values in time. We have tried to overcome it. We will discuss the
possible sources of the divergence of results in section \ref{coordination}.

The measurement of temperature dependence of $\alpha $ was rarely
attempted, since the large divergence of obtained $\alpha $ values
obscures the effect. Two recent studies by Li et al. \cite{Li} and
by Winkler et al. \cite{Winkler} can serve as an example. The
authors of the first study (Boston College/Aerodyne Research Inc.
group) found that $\alpha $ decreases with temperature within the
temperature range between 257 and 280 K. The authors of the second
study (University of Vienna/University of Helsinki group) claim
that $\alpha $ and $\alpha _{T}$ exhibit no temperature dependence
between 250 and 290 K. The comparison of these results can be
found in \cite{4x}.

In this paper we present our new experimental results of evaporation coefficient of
water in air, as well as our reprocessed results for water in nitrogen (compare
\cite{vsTemp}), versus temperature; both under atmospheric pressure. The results for
air and for nitrogen are consistent, which also indicates that the presence of small
amounts of such soluble and/or reactant gases as CO$_2$ in the ambient air, does not
influence the value of kinetic coefficients. In comparison to our previous works we
refined our data processing which enabled us to determine the droplet radius with
higher accuracy and trace its evolution with higher confidence. Smoother radius
derivatives enabled us, in turn, to employ direct fitting of the model in finding
the kinetic coefficients and avoid most approximations. We also operated on a larger
set of experimental runs. This yielded correction of the values of kinetic
coefficients we obtained previously, and revealed different temperature dependence.
These results turned out to be in excellent agreement with the results of BC/ARI
group: the values of $\alpha $ coincide within the temperature range from $\sim 273$
to $\sim 281$ K and our results extending towards higher temperature follow the
temperature dependence found by BC/ARI group. Since BC/ARI group and our results
together span over larger temperature range, the accuracy of finding temperature
dependence of $\alpha $ could be improved.

\section{Experiment}

\begin{figure}[htb]
\begin{center}
\scalebox{0.50}{\includegraphics{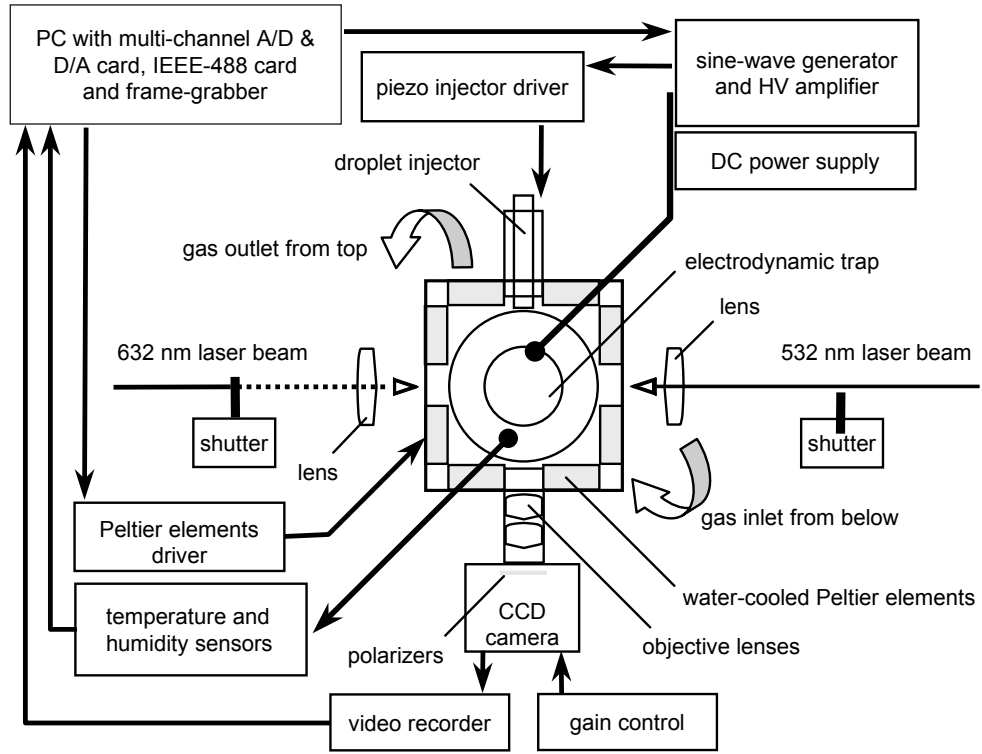}}
\end{center}
\caption{Experimental setup - top view.} \label{setup}
\end{figure}

The experimental setup is presented in figure \ref{setup}. It consisted of a
hyperboloidal electrodynamic quadrupole trap (see e.g. \cite{Major}), kept in a
small climatic chamber. The high resistivity electric circuit of the trap drive
enabled us to operate in a humid atmosphere. A detailed description of this
apparatus can be found in \cite{vsTemp,Jakubczyk} and of further modifications in
\cite{IOPD,JOSAA}.

The droplets were introduced into the trap with a piezo injector, similar to
constructed e.g. by Lee et al. \cite{lee} or Zoltan \cite{zoltan}. The injection
timing was controlled relative to the trap driving AC signal. By choosing the proper
injection phase, the sign and, to a certain extent, also the value of the charge of
the injected droplet could be controlled. The initial temperature of the droplet was
that of the chamber.

In each experiment, the chamber was first flushed with clean, dry nitrogen, and then
filled with a mixture of nitrogen/air and water vapor. The temperature in the
chamber was monitored and stabilized. A zone type temperature control enabled us to
eliminate vertical temperature gradients. Horizontal gradients were found
negligible.

The humidity in the chamber, but outside of the trap was monitored continuously with
semiconductor sensors. Due to poor vapor exchange through trap openings accompanied
by injecting liquid water into the trap, the humidity inside the trap could not be
inferred directly from those measurements. The value of humidity in the trap found
as fitting parameter (see section \ref{processing}) turned out to be higher by
several percent than sensors readings. Resorting to fitting method was inevitable,
since the humidity accuracy required for the correct assessment of kinetic
coefficients was inaccessible via any on-line sensor measurement. On the contrary,
analyzing droplet radius evolution seems a highly accurate method of measuring
relative humidity, surpassing any on-line methods.

In our experiments we used ultra-pure water. The details about its initial
parameters and sample preparation can be found in \cite{vsTemp}, where we discussed
also the absorption of impurities by ultra-pure water, and their influence upon the
experimental results there.

Droplet evolutions were studied with time resolved static light
scattering, with green or red laser light. We found no
inconsistency between the results obtained for both and we infer
that the light wavelength had no influence upon the results.

\section{Evaporation model}

In order to interpret the experimental results, a model of
evaporation was necessary. The model of evaporation we used was
based on a generally accepted model which can be found in such
textbooks as \cite{Pruppacher,Fuchs,Friedlander}. It was a
slightly rephrased and numerically reexamined version of what we
had used previously \cite{vsTemp,mz_water}. Below we discuss the
details of the model equations we used, since the results may
depend significantly on the apparently minute approximations made.
We also point to a certain approximation typically made that we
found weighing heavily upon the results.

The quasi-stationary evaporation of a free, motionless droplet
larger than the mean free path of vapor molecules, can be easily
described with the diffusion equations with boundary conditions
defined by the thermodynamic conditions in the reservoir (far from
the droplet). This part of the model does not rise much
difficulties as long as the characteristic times of the process
justify the quasi-stationary approach \cite{Worsnop,Schwartz}.

For small droplets, of the size comparable with the mean free path
of vapor molecules, the language of diffusion is not appropriate.
The transport of mass and heat below the mean free path distance
from the surface must be perceived as effusion or evaporation into
vacuum. The net effusive flow of vapor can be expressed as
difference between outgoing and incoming effusive flows
\cite{Pruppacher,Present}:
\begin{equation}
J=\pi a^2 \alpha \left[ \rho _e(r=a)\overline{v}(T_a)-\rho
(r=a+\Delta )\overline{v}(T_{a+\Delta})\right] \mbox{ ,}
\label{vacuum}
\end{equation}
where $\overline{v}(T)=\sqrt{8RT/\pi M}$ is the mean absolute
thermal velocity of vapor molecules for the temperature $T$; $T_a$
is the temperature of the droplet (surface), $T_{a+\Delta}$ is the
temperature of vapor at the distance $\Delta$ (comparable with the
mean free path of the vapor molecule) from the surface. $\rho (r)$
is the vapor density at the distance $r$ from the droplet center
while $\rho _e(r=a)$ is the vapor density at the droplet surface
for the equilibrium conditions (steady state, no net flow).

The usual approximation made is $T_a=T_{a+\Delta}$ (see
e.g.\cite{Pruppacher}). It implies neglecting the slowing down of
the mass transport by thermal effusion (a short range analogue of
thermal diffusion). It should also be noted, that lifting the
temperature dependence of $\overline{v}$ introduces some
additional temperature dependence into $\alpha $. Unfortunately,
discarding this usual approximation excludes using standard
solutions and substantially complicates calculations. To overcome
such difficulties, we decided to introduce a simple correction of
$\alpha$ at the end. We shall address this issue in detail later.
Following the standard route, we compare effusive and diffusive
flows at $r=a+\Delta$. Since these flows are equal and both are
proportional to the vapor density gradient it is possible to write
a compact expression:
\begin{eqnarray}
a\frac{da}{dt}& = & \frac{MD_k(a,T_{a})}{R\rho _L}\left[
S\frac{p_{\infty }(T_R )}{T_R }-\frac{p_a(T_{a})}{T_a }\right]= \label{basic}\\
& = & \frac{MD_k(a,T_{a})}{R\rho _L} \frac{p_{\infty }(T_R )}{T_R
} \left[ S-\frac{p_a(T_{a})}{p_{\infty }(T_a )} \frac{p_{\infty
}(T_a )}{p_{\infty }(T_R )}
\frac{T_R }{T_{a}} \right]\mbox{,} \nonumber \\
\end{eqnarray}
where
\begin{equation}
\frac{p_a(T_{a})}{p_{\infty }(T_a )}=\exp \left[ \frac M{RT_{a}
\rho _L}\left( \frac{ 2\gamma }a-\frac{Q^2}{32\pi ^2\varepsilon
_0a^4}\right) \right]
\end{equation}
is the Kelvin equation, accounting for the modification of
equilibrium vapor density near the droplet surface due to the
surface curvature and charge effects \cite{Friedlander}, and
\begin{equation}
\frac{p_{\infty }(T_{a})}{p_{\infty }(T_R )}=\exp \left[ \frac
{qM}{R}\left( \frac{1}{T_R}-\frac{1}{T_a}\right) \right]
\end{equation}
is the Clausius-Clapeyron equation. The effective diffusion
coefficient $D_k$ accounts for the effect of effusion:
\begin{equation}
D_k=\frac D{a/(a+\Delta _C )+D\sqrt{2\pi M/(RT_a)}/(a\alpha )}
\mbox{ .}  \label{Dk}
\end{equation}
$D$ is the diffusion constant for water vapor in nitrogen/air,
$T_R $ is the temperature of the reservoir, $Q$ is the droplet
charge, $p_{\infty }$ and $p_a$ are the equilibrium (saturated)
vapor pressure above the flat interface and above the interface of
the curvature radius $a$ at a given temperature. $S$ is relative
humidity. $\gamma $, $\rho _L$, $M$ and $q$ are the surface
tension, density, molecular mass and the latent heat of
evaporation of liquid water, $\varepsilon _0$ is the permittivity
of vacuum, $R$ is the universal gas constant. $\Delta _C$ defines
the effective range of the gas kinetic effects. It is comparable
with the mean free path of particles of surrounding gaseous medium
$\lambda _a$. We assumed $\Delta _C=4\lambda _a/3$
\cite{Pruppacher}.

The change of droplet mass by evaporation (condensation) is
associated with heat absorption (release), which manifests as
temperature drop (rise) toward the droplet. The equation for the
transport of heat can be presented in a convenient form:
\begin{equation}
a\frac{da}{dt} = \frac{\lambda _K(a,T_{a})}{q \rho _L}(T_{a}-T_R)
\mbox{ ,} \label{dT}
\end{equation}
where
\begin{equation}
\lambda _K=\frac \lambda {a/(a+\Delta _T)+ \lambda \sqrt{2\pi
M_N/(RT_a)}/(a\alpha _T\rho _Nc_P)}\mbox{ ,} \label{Lk}
\end{equation}
is the effective thermal conductivity of moist nitrogen (air) and
$\lambda $, $\rho _N$, $M_N$ and $c_P$ are thermal conductivity,
density, molecular mass and specific heat capacity under constant
pressure of moist air/nitrogen respectively.  $\Delta _T$ plays
role analogous to $\Delta _C$ and was assumed as $\Delta _T=\Delta
_C+4\lambda /(\tilde{v}c_P\rho _N)$. Since in the vicinity of
standard temperature and pressure, the partial pressure of water
vapor can be neglected in comparison to that of air/nitrogen, it
can be assumed that the heat is conducted to the droplet mostly by
the molecules of air/nitrogen. In consequence, the flux of mass
can be considered independently of the flux of heat and $\Delta
_C$ associated with the transport of mass should be distinguished
from $\Delta _T$ associated with the transport of heat.

The direct influence of the droplet charge, through charge-dipole interaction, upon
the mass (or heat) transport was estimated to be negligible for droplet charge
ensuring Coulomb stability (compare \cite{Nadykto}). Similarly, field emission did
not take place for surface charge densities encountered in our experiments (see e.g.
\cite{Castano}). The Coulomb explosion of the droplet is a threshold process and
does not need accounting in the transport equations.

\section{Experimental data processing} \label{processing}

The procedure of the numerical processing of experimental data,
which we found the most stable and yielding the most consistent
results, relies on the direct fit of the model equations to the
experimentally obtained droplet radius change rate $\dot{a}\equiv
da/dt$ as a function of droplet radius $a$. The data preparation
procedure is presented below.

\begin{figure}[htb]
\scalebox{0.46}{\includegraphics{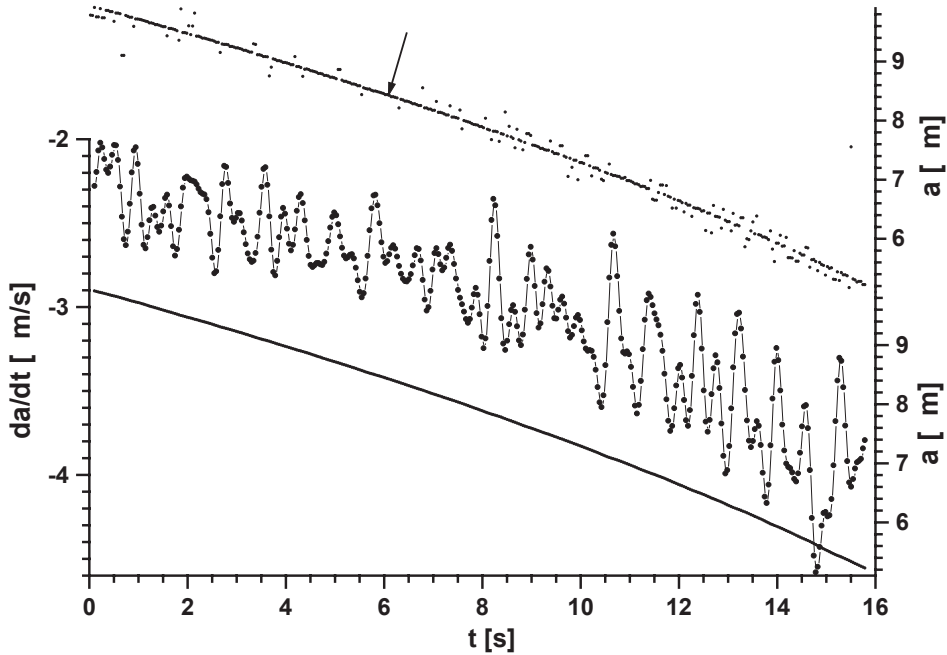}} \caption{An example
of temporal droplet radius evolution, before and after processing -
top and bottom curve respectively. Derivative calculated from
processed data - middle curve. $N=395$, $T_R=283$~K,
$p_{atm}=1006$~hPa, $S_{sens}=0.9$.} \label{a_a_dadt}
\end{figure}

\begin{figure}[htb]
\scalebox{0.46}{\includegraphics{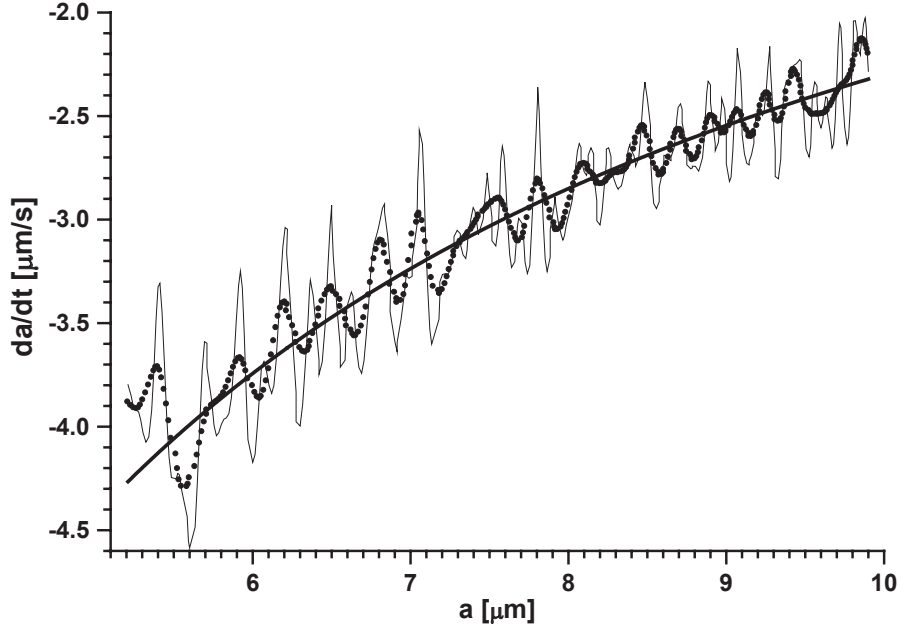}} \caption{Droplet
radius temporal derivative versus droplet radius, corresponding to
figure \ref{a_a_dadt}, before and after filtering (points). The
result of model fitting - solid line. Fitting parameters:
$S_{fit}=0.9762$, $Q=3.7\times 10^6$ elementary charge units,
$\alpha =0.155$, $\alpha _T=1$.} \label{dadt_a}
\end{figure}

The running radius of the droplet $a_i(t_i)$ was obtained (off line) from the
angularly resolved Mie scattering pattern for the time $t_i$ with the help of a
gradientless fitting procedure ("library method"). Each droplet evolution yielded a
sequence of a few hundreds data points indexed with $i$ (see figure \ref{a_a_dadt}).
We had found that in order to obtain reliable results, significant care must be
taken to ensure a high signal to noise ratio of the measurement. There happen data
points misplaced to incorrect "evolution branch", associated with the Mie resonances
that could not be handled with the method used (see description of the method
\cite{vsTemp}). The accuracy of a single value of droplet radius $a_i$ (except for
misplaced points) was estimated as $\pm 15$~nm. The $a_i(t_i)$ sequence was stripped
to main "evolution branch" (indicated with arrow in figure \ref{a_a_dadt})and
interpolated in order to obtain regularly spaced data points. The time derivative
$\dot{a_i}(t_i)$ was calculated (figure \ref{a_a_dadt}). The $a_i(t_i)$ evolution
was smoothed with low pass FFT filter and combined with the derivative in order to
obtain $\dot{a_i}(a_i)$. Finally $\dot{a_i}(a_i)$ was smoothed (figure
\ref{dadt_a}).

On the other hand, subtracting equation \ref{basic} from equation
\ref{dT} leads to an equation binding $T_a$ and $a$. For every
experimental $a_i$ this equation can be unambiguously numerically
solved for $T_a$, yielding $T_a(a_i)$. This, on insertion into
equation \ref{basic} yields at every experimental data point a
numerically solvable equation binding $\dot{a}$ and $a_i$. Thus, a
model prediction of $\dot{a}(a_i)$ could be obtained.

In order to find $\alpha $, $\alpha _T$, $S$ and $Q$, we minimized
the function
\begin{equation}
\chi ^2=\frac{\chi _0^2}{N}\sum ^N_{i=1} \left[
\dot{a_i}(t_i)-\dot{a}\left( a_i(t_i),\alpha ,\alpha _T, S,Q
\right) \right]^2
\end{equation}
using a gradient method. $N$ was the total number of experimental data points of the
evolution, and $\chi _0$ was an arbitrarily chosen normalizing factor. $\alpha $ and
$S$ were found to be the essential parameters and could be unambiguously determined,
while $\alpha _T$ and $Q$ could be determined only with limited confidence. Since
$\alpha $ and $S$ had seemed partially interconnected, the minimization was
performed very carefully, starting from various combinations of $\alpha $ and $S$
($\alpha $ larger, $S$ smaller versus $\alpha $ smaller $S$ larger) and accepted
only if leading to the same results. The less relevant parameters were initialized
as follows: $\alpha _T=1$, (values above 1 were allowed; compare \cite{Winkler}),
and $Q=8\pi \sqrt{ \varepsilon _{0}\gamma a_i^{3}}$, where $a_i$ corresponded to the
smallest droplet radius observed in the evolution (no Coulomb instabilities during
evolution). The resulting $Q$ was much approximate, and we couldn't detect the
eventual droplet charge loss (see eg. \cite{Duft}) by analyzing the evolution of the
droplet radius. The minimization was also hardly sensitive to $\alpha _T$, however a
value close to unity could be inferred. Since for larger droplets ($a>6$~$\mu$m) the
kinetic effects as well as the effect of the droplet charge were negligible, only
$S$ was fitted in this range, as a first step, and then the minimization was
extended towards smaller radii with $\alpha $ added as a parameter. Finally $\alpha
_T$ and $Q$ parameters were added. The whole procedure exhibited best stability for
$S>95\%$, since the evaporation was slower then (compare equation \ref{basic}) and
thus: (i) the evolution of the droplet radius could be determined with high
precision and (ii) the temperature jump at the interface $\Delta T$ was so small
(compare equation \ref{dT}) that the model equations used were exact enough. It
would be valuable to validate the procedure of finding kinetic coefficients using
other liquids (such as ethylene glycol). Unfortunately, the parameters such as
diffusion constant are usually not known with adequate precision. On the other hand,
after slight modification of the procedure, it should be possible to look just for
diffusion constant, which we intend to do soon.

\subsection{Correction of \protect$\alpha$}

In order to estimate the influence of $\Delta T$ upon the obtained value of $\alpha
$, we apply an approximation $T_{a+\Delta}=T_R$ to formula \ref{vacuum}, which is
opposite to usually applied $\Delta T=0$, and we compare the results of both
approximations. The approximation that we introduce means that we account only for
thermal effusion while neglecting thermal diffusion. Since for our experimental
conditions the temperature gradient was highest in the very vicinity of the
interface (see \cite{Fang}), our approximation was legitimate. For simplicity we
also assumed that the shape of distribution of vapor density was spatially constant
and temperature independent. It implied $\rho (r=a+\Delta )=S_{a+\Delta}\rho (T_R)$,
where $S_{a+\Delta}=const$ represented relative humidity at $r=a+\Delta$. If we
require that the effusive flows calculated with each of the approximations are
equal, we have:
\begin{equation}
\frac{\alpha }{\alpha (\Delta
T=0)}=\frac{S_{a+\Delta}-\frac{\rho_e(T_a )}{\rho_e(T_R
)}\sqrt{\frac{T_R}{T_a}}}{S_{a+\Delta}-1}\simeq
\frac{S_{a+\Delta}-\frac{\rho_e(T_a )}{\rho_e(T_R
)}}{S_{a+\Delta}-1} \mbox{ .} \label{correction}
\end{equation}
Introducing $T_a(a_i)$ (see section \ref{processing}) into equation
\ref{correction}, we can find a correction of $\alpha $, where $S_{a+\Delta}$ is a
(scaling) parameter. It is initiated as $S_{a+\Delta}=S$ and optimized so that
$\alpha /\alpha (\Delta T=0)\rightarrow 1$ for $\Delta T\rightarrow 0$ (larger $T_a$
in case of our experiment; see the inset in figure \ref{rawresults}). The results
presented in figure \ref{alfasvsTa} are already corrected. In our case a significant
(by a factor of nearly 2) correction was near the freezing point and by several
percent at 276.5 K. The equation \ref{correction} is essentially approximate and
leads to underestimation of $\alpha $. It can be seen in figure \ref{alfasvsTa} -
our data points seem to lie slightly below the trend line. It turns out that for
many reasonable experimental conditions the correction factor can be higher than 2.
We shall discuss a few examples in section \ref{coordination}. Considering the
approximations made, we estimate that for thermodynamic conditions encountered in
atmosphere the accuracy of the correction factor should not be worse than several
percent.

\section{Results and Discussion}

\begin{figure}[htb]
\scalebox{0.46}{\includegraphics{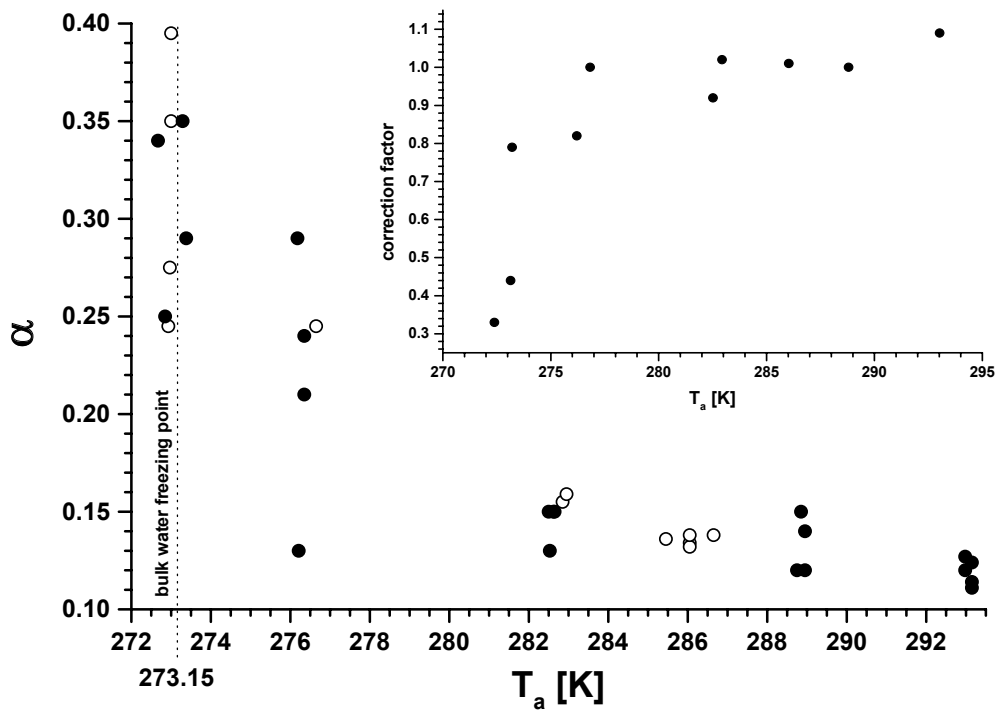}} \caption{Non-averaged
experimentally obtained values of $\alpha $ as a function of droplet
surface temperature. Solid and open circles represent results
obtained for nitrogen and air respectively. The corresponding
calculated evaporation coefficient correction factors, due to the
thermal effusion, are presented in the inset.} \label{rawresults}
\end{figure}

The raw results are presented in figure \ref{rawresults} as a
function of the droplet (surface) temperature. The kinetic
coefficients should be presented as a function of the droplet
(surface) temperature, since in general, due to evaporative
cooling, it may differ significantly from the temperature of the
reservoir. In case of BC/ARI group experiments, $T_R-T_a\leq2$~K
\cite{Worsnop}. Though, in our case $T_R-T_a\leq0.7$~K only, it is
sufficient that some of our results correspond to supercooled
water as well.

The kinetic coefficients found for water droplets in nitrogen and
in air were mutually compatible (see figure \ref{rawresults}). It
implies, that the gases absorbed by water from the air had
negligible impact upon our measurements and generally there is no
strong dependence upon the composition of the ambient atmosphere.

\begin{figure}[htb]
\scalebox{0.46}{\includegraphics{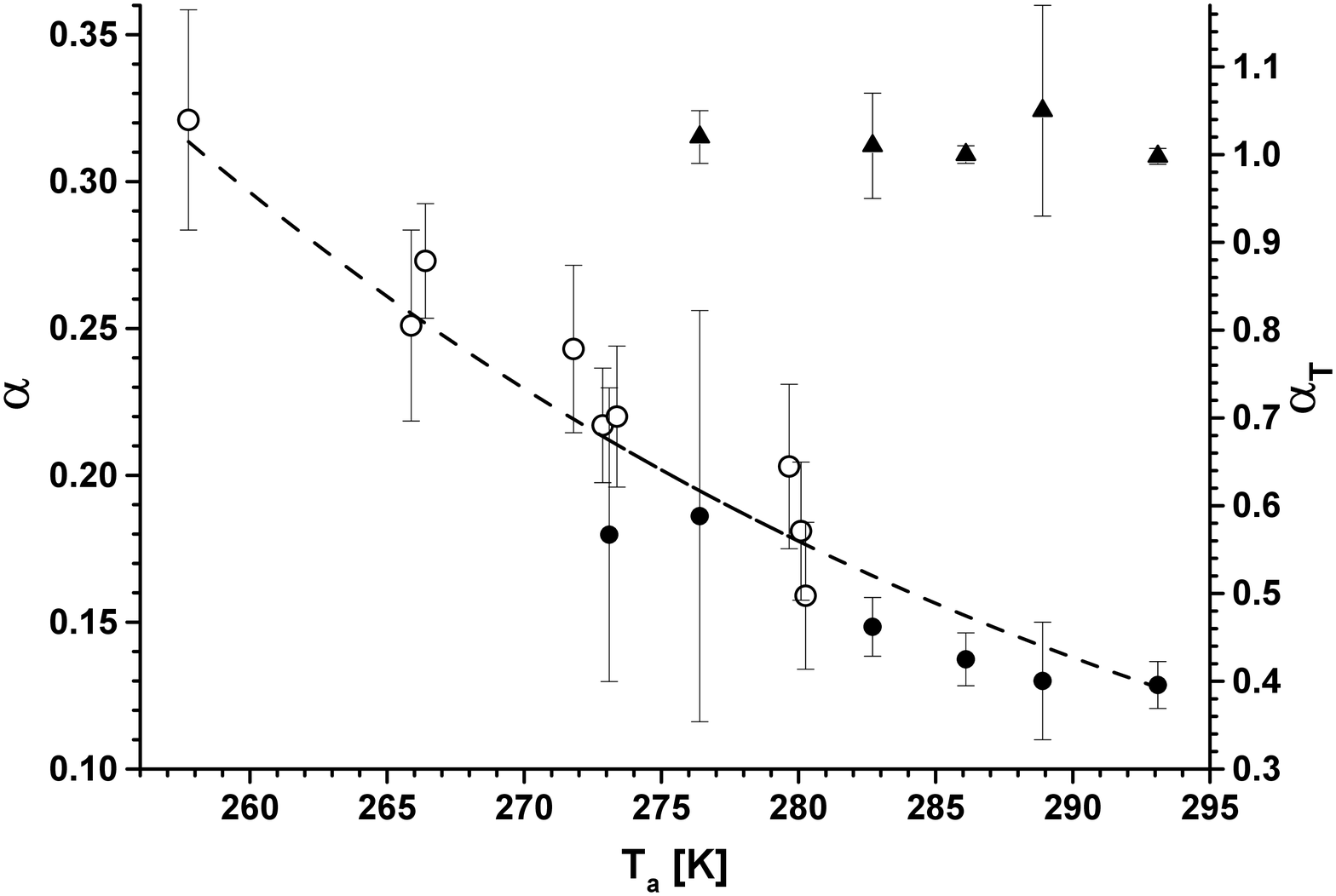}} \caption{Collected
$\alpha $ and $\alpha _T$ values as a function of droplet surface
temperature. Solid circles and triangles: corrected evaporation
coefficient and thermal accommodation coefficient respectively,
obtained from our measurements; hollow circles: condensation
coefficient measured by BC/ARI group \cite{Li}. Dashed line
represents the fit of the equation \ref{TST} to the results of
BC/ARI group and our data together.} \label{alfasvsTa}
\end{figure}

The final results are shown in figure \ref{alfasvsTa}. There are
values of evaporation coefficient we obtained (solid circles) and
values obtained by BC/ARI group, taken from \cite{Li} (hollow
circles). The values of thermal accommodation coefficient we
obtained are also presented (solid squares). Data points
corresponding to our results were obtained by averaging the raw
results (compare figure \ref{rawresults}). We also followed BC/ARI
group and used the formula they derived basing on Transition State
Theory (TST) (e.g. equation 7 in \cite{Nathanson}). Such
formulation enables expressing the results in the language of
thermodynamic potentials:
\begin{equation}
\frac{\alpha }{1-\alpha }=exp\left( \Delta G_{obs}\right)\mbox{ ,}
\label{TST}
\end{equation}
where $\Delta G_{obs}$ is the Gibbs free energy and its temperature dependence can
be expressed as $\Delta G_{obs}=\Delta H_{obs}-T\Delta S_{obs}$. $\Delta H_{obs}$
and $T\Delta S_{obs}$ are treated just as parameters, their physical meaning is not
clear (see discussion below). This formula is derived on an assumption, well
justified with elegant experiments by Nathanson et al. described in
\cite{Nathanson}, that the particles from the gaseous phase enter liquid via an
intermediate surface state. Dashed line in figure \ref{alfasvsTa} represents the fit
we made to the results of BC/ARI group and our data points together. It yielded
$\Delta H_{obs}=4830\pm 150$~cal/mol and $\Delta S_{obs}=20.3\pm 0.5$~cal/mol, which
is within the limits of error equal to the values given in \cite{Li}, i.e. $\Delta
H_{obs}=4.8\pm 0.5$~kcal/mol and $\Delta S_{obs}=20.3\pm 1.8$~cal/mol. The accuracy
of our fit (and so of the values obtained) is higher due to the larger number of
data points.

The comparison of our results with those of BC/ARI group indicates
also that there was no perceivable influence of droplet charge
upon kinetic coefficients. Vibrating orifice injector generates,
at least in average, neutral droplets, while in our experiments
with evaporating charged droplets it could be assumed that the
charge was approaching its maximum value - the Rayleigh limit.
Similarly, the comparison of aforementioned experiments reveals no
measurable influence of ambient atmosphere pressure upon the value
of kinetic coefficients.

The temperature dependence of $\alpha $, though obtained with
essentially different method, coincides with the results of BC/ARI
group (see eg. \cite{4x}). Our result extends into higher
temperature range. Furthermore, we measured evaporation
coefficient while BC/ARI group measured condensation coefficient.
It supports the notion of equivalence of these coefficients.

The thermal accommodation coefficient we obtained $\alpha
_{T}=1\pm 0.05$ (figure \ref{alfasvsTa}) agrees with both BC/ARI
and UV/UH groups' results. However it is hard to asses the real
uncertainty of $\alpha _{T}$; the statistical error we found may
be too small (see section \ref{processing}). Thus it is not
possible to derive information on its temperature dependence.
Recently, there seems to arise a general consensus that $\alpha
_{T}$ is close to 1, which means, that all the particles striking
the interface thermalize.

\subsection{An attempt of results coordination \label{coordination}}

Since it is quite improbable that all the kinetic coefficients
measurements performed over the years were loaded with random
errors, it must be assumed that the experiments, though accurate
by themselves, measured different quantities. Many authors have
tried to coordinate the results by pointing out what was really
measured (see e.g.: \cite{Pruppacher,Marek}). However there is no
consensus. We shall also try to address this issue.

The divergence of results obtained by different authors has been usually attributed
to: (i) difficulties in accounting for various physical and chemical interfacial
processes; (ii) effects of impurities, and especially surface active agents
\cite{Feingold}; (iii) structure of the interface (dynamic surface tension, reaching
the balance by the interface) and (iv) dependence of the coefficient value upon the
model used (indirectness of measurement). It has been pointed out
\cite{Pruppacher,Marek} that two classes of experiments could be distinguished: (i)
with a quasi-static interface, yielding $\alpha <0.1$ and (ii) with a continuously
renewing surface, yielding $\alpha \geq0.1$. However, such categorization requires
defining the time scale. Such scale has not been agreed yet, neither the leading
mechanism responsible for interface aging. For example, the characteristic times
used in Molecular Dynamics (MD) studies are only hundreds of ps. This falls into a
non-stationary interval, when the transients in the temperature and vapor density
fields are starting to form. The Transition State Theory (TST) considerations of
Nagayama et al. \cite{Nagayama}, seem to be in agreement with MD calculations and
predict $\alpha \simeq 1$ around room temperature. However, it is worth noting,
that, for example, stationary values of the surface tension are reached within
milliseconds \cite{Marek} and all these time scales are far below the characteristic
timescale of cloud droplet growth process, which lie in the range of seconds (or
even minutes) \cite{Chuang}.

Recently, Fukuta and Myers \cite{FukutaMyers} have noticed, that accounting for the
effect of moving gas-liquid interface ("moving boundary effect") can change the
resulting value of kinetic coefficients by several percent. In their work they
managed to account for this effect in an elegant way. Though the thermodynamical
conditions and the velocity of the interface in our experiment were similar to
theirs, in present work, we have decided to neglect the moving boundary effect,
since the correction of mass accommodation coefficient we propose is much larger.

In this paper we would like to point to a mechanism which falls within the 4th
category - model dependent, however it is related to the issue of the characteristic
timescale of the process and its distance from the thermodynamic equilibrium.
Usually, authors are careful to estimate the characteristic times of mass and heat
transport processes involved, in order to assure the proper description. It seems,
that in some cases this alone can be somewhat misleading, because of the thermal
effusion which we already mentioned. We shall consider four examples.

In case of BC/ARI group experiments, the vapor-liquid contact
lasts several milliseconds but the droplet is essentially in
equilibrium with the reservoir. In order to achieve temperatures
below 273 K the evaporative cooling was used which inevitably
caused temperature jump near the surface (up to 2 K) and thermal
effusion as a consequence. However, since the value of $\alpha$
was not obtained from the evolution of droplet radius, its value
should be safe and no correction is needed.

In our case, we selected for the analysis the droplet evolutions
which lasted a few seconds which guaranteed that the process had
been quasi-stationary in the diffusion time scale. For faster
evolutions the temperature jump approached 1 K, and since $\alpha$
was obtained from the evolution, it had to be corrected by means
presented above.

In case of the experiment of UV/UH group \cite{Winkler}, the evolution lasted $\sim
50$ ms, which is shorter than in our case, but for the thermodynamic conditions they
had, the process still could be regarded as quasi-stationary. However, the
temperature jump of $\sim 3$ K could be expected for such evolution. This alone
would require a correction of $\alpha $ by a factor of 2. Further overestimation
might be caused by uncertainty of water vapor saturation. There are also rather few
data points lying on a relatively flat curve, which as we know from our experience,
causes the increase of measurement uncertainty.

Lastly, in case of very interesting Fukuta and Myers experiment \cite{FukutaMyers}
the evolution (condensation) lasted $\sim 3$~s (similarly as in our experiment).
Since the final droplet radius was $\sim 2$~$\mu $m, it can be inferred that
$\dot{a}\approx 1$~$\mu $m/s, which in turn yields temperature jump of only $\sim
0.2$ K. However, since the mass transport was relatively slow (supersaturation used
was very small), the effect of even small temperature jump at the interface could be
relatively large. According to our estimation (see expression \ref{correction}) the
correction of mass accommodation coefficient should be as high as 5! This would
bring Fukuta and Myers result for NaCl and (NH$_4$)$_2$SO$_4$ at 277~K to $\alpha
\simeq 0.2$, which agrees within the limits of error with ours and BC/ARI group
results, even allowing for moving boundary effect which we neglected.

\section{Conclusions}

We conclude that it is feasible to obtain reliable values of
evaporation coefficient by analyzing the evaporation of a small
droplet. It requires however several tens of data points per
evolution and droplet radius measurement accuracy of several
nanometers. Generally accepted model of quasi-stationary
evaporation seems sufficient for experimental data analysis in
most cases. We found however that when evaporative cooling of the
droplet becomes of the order of 1 K, it is necessary to consider
the effect of thermal effusion, which is a short distance analogue
of thermal diffusion. The kinetic coefficients found for water
droplets in nitrogen and in air were mutually compatible. The
evaporation coefficient for the temperature range from 293.1 K
down to 273.1 K was found to increase from $\sim 0.13$ to $\sim
0.18$ and follow the trend given by Arrhenius formula (see
\ref{TST}) with the parameters $\Delta H_{obs}=4830\pm
150$~cal/mol and $\Delta S_{obs}=20.3\pm 0.5$~cal/mol. This
temperature dependence is in excellent agreement with the results
of BC/ARI group, which concern condensation coefficient, were
obtained with essentially different technique for much lower
ambient gas pressure and extend toward lower temperatures. The
comparison with BC/ARI group experiments enables to draw a few
additional conclusions: (i) the evaporation and condensation
coefficients are essentially equivalent; (ii) there was no
measurable influence of ambient atmosphere pressure upon the value
of kinetic coefficients in the range from $\sim $~kPa to $\sim
100$~kPa; (iii) there was no measurable influence of droplet
charge upon the value of kinetic coefficients up to the Rayleigh
stability limit. The value of thermal accommodation coefficient we
obtained $\alpha _{T}=1\pm 0.05$ agrees well with recent results
of many authors.

{\bf Acknowledgment.} This work was supported by Polish Ministry
of Science and Higher Education grant No.~1~P03B~117~29.

\end{document}